# Aligning with Human Values to Enhance Interaction: An eHMI-Mediated Lane-Changing Negotiation Strategy Using Bayesian Inference

Boyao Peng
*Department of Traffic Engineering & Key Laboratory of Road and Traffic Engineering of Ministry of Education, Tongji University*
Shanghai, China
Peng_3U@163.com

Linkun Liu
*Department of Traffic Engineering & Key Laboratory of Road and Traffic Engineering of Ministry of Education, Tongji University*
Shanghai, China
aliulinkun@tongji.edu.cn

*Abstract*—As autonomous driving technology evolves, ensuring the stability and safety of Autonomous Driving Systems (ADS) through alignment with human values becomes increasingly crucial. While existing research emphasizes the adherence of AI to honest ethical principles, it overlooks the potential benefits of benevolent deception, which maximize overall payoffs. This study proposes a game-theoretic model for lane-changing scenarios, incorporating Bayesian inference to capture dynamic changes in human trust during interactions under external Human-Machine Interface (eHMI) disclosed information. Case studies reveal that benevolent deception can enhance the efficiency of interaction in up to 59.4% of scenarios and improve safety in up to 52.7%. However, in the most pronounced cases, deception also led to trust collapse in up to 36.9% of drivers, exposing a critical vulnerability in the ethical design of ADS. The findings suggest that aligning ADS with comprehensive human ethical values, including the conditional use of benevolent deception, can enhance human-machine interaction. Additionally, the risk of trust collapse remains a major ethical loophole that must be addressed in future ADS development.

## I. Introduction

Autonomous driving technology is advancing at a rapid pace, with Highly Automated Vehicles (HAVs, Level L3 and above) considered capable of making independent decisions in complex scenarios. Nevertheless, their market share is expected to reach only around 40% by 2035 [1], indicating that HAVs will continue to coexist with human-driven vehicles (HVs) for the foreseeable future. This mixed traffic environment poses significant challenges, as current technology still struggles to ensure safe and reliable interaction between HAVs and HVs.

The alignment problem is posing a critical threat to the reliability of HAVs' interaction with HVs, as it may lead to unsafe or unpredictable behavior. Alignment problem [2], [3] is a concept in artificial intelligence (AI). It refers to the situation where AI systems deviate from human-intended goals, preferences, or ethical principles. Such deviations may cause AI to achieve its objectives in unexpected ways and harm humanity. According to the RICE principle [3], an AI system that is aligned with human values should be controllable and interpretable, pursuing the human pre-specified objectives, while adhering to socially accepted ethical standards and values in its decision-making. However, human ethical systems exhibit intrinsic complexity without a universally accepted standard [4]. In certain situations, people may adopt ethically controversial decisions in order to achieve their intended goals. A typical example is "benevolent deception", which is not driven by malicious intent but rather to advance altruistic objectives. For example, doctors will keep their patients optimistic through well-intentioned lies to help with the disease. This consideration raises an important ethical question. Benevolent deception is a part of the human ethical system, yet the act of deception itself is often regarded as morally questionable. This raises the issue of whether AI systems should be aligned with such complex and potentially contradictory aspects of human values.

Given that HAVs primarily rely on automated driving systems (ADS), which are a form of AI system, they also face the alignment problem. Specifically, ensuring that HAVs' decision-making aligns with human values and safety standards is crucial for their safe interaction with HVs. Currently, ADS is often trained using machine learning methods such as end-to-end learning and deep reinforcement learning [5], [6]. While these methods enable the system to learn human behavioral patterns, they do not inherently incorporate human ethical values into decision-making. Additionally, the lack of interpretability in these approaches makes it difficult to verify the system's decisions, especially when they diverge from ethical principles. More seriously, any subsequent technological advancement will struggle to establish trustworthy interaction benchmarks when the ADS is unable to align with human values.

Similarly, HAVs may also conduct benevolent deception to optimize traffic efficiency or reduce collision probabilities. Such practice of deception for global optimization reveals the conflict between goal alignment (executing predefined operational targets) and value alignment (embodying deontological ethical constraints). This tension prompts critical inquiry along two dimensions: whether ADS can achieve technical alignment with the context-sensitive human moral framework, and whether such alignment can sustain operational trustworthiness amid the inherent ethical indeterminacy of value-laden decision-making.

The external Human-Machine Interface (eHMI) carried by HAVs is an important medium for interaction with other human drivers. It uses animations, patterns, or text to communicate driving status and intentions to external traffic participants. Fig. 1 presents several examples of eHMIs, such as the Mercedes-Benz F 015 Luxury in Motion [7], which projects dynamic curved lines and a moving zebra crossing to signal yielding to pedestrians. eHMI holds potential for facilitating human-machine cooperation, provided that value alignment with humans is achieved. Existing research lacks consideration of the information disclosure mechanisms of eHMI and tends to tacit assumption that the eHMI will disclose the behavioral intentions with full honesty.

Liu et al. [8] have conducted a preliminary study on benevolent deception in unprotected left-turn interactions. Their work showed that benevolent deception improves safety in single-shot decisions. However, unprotected left-turn interactions are too brief to reflect dynamic strategy changes. And they did not consider the heterogeneity of HVs, which leads to varying levels of trust in HAVs, especially when the vehicle implements benevolent deception. In lane changing interactions, continuous decision-making and frequent adjustments are required, which makes the effectiveness of benevolent deception unclear. Unlike unprotected left-turn interactions, lane changing involves not only interactions between two vehicles but also requires the consideration of other vehicles' states. Furthermore, since lane changing is common on highways and urban expressways — where vehicles typically travel at higher speeds — collisions in such scenarios are more likely to result in severe consequences. Therefore, our work focused on lane change scenarios, exploring the impact of benevolent deception in lane changing interactions.

We investigated the implementation of benevolent deception through eHMI, exploring the feasibility of achieving enhanced human-machine cooperation and value alignment through strategic information manipulation while preserving human trust. A three-stage methodology is presented in this study. First, we developed a dynamic Bayesian trust estimation framework that classifies HVs into cooperative or non-cooperative styles based on their behavior, addressing heterogeneity. This classification is then used to estimate trust, as shown in Fig. 2(a). Second, a dynamic game model for lane changing was created, which integrates trust estimation into the payoff function and considers the influence of surrounding vehicles, as shown in Fig. 2(b). Finally, numerical experiments validated the model and optimized the strategy for information disclosure, balancing cooperation and trust preservation. Efficiency is measured by interaction duration, and safety by collision risk.

The main contributions of this paper are threefold: (1) A dynamic estimation framework for human trustworthiness is proposed, enabling the quantification and computation of human trust in eHMI. (2) Introduced strategic dishonesty as part of the alignment objective, demonstrating its potential to enhance human-machine cooperation without breaching trust boundaries. (3) Proposed an optimal eHMI disclosure strategy that balances trust preservation with improved interaction payoff.

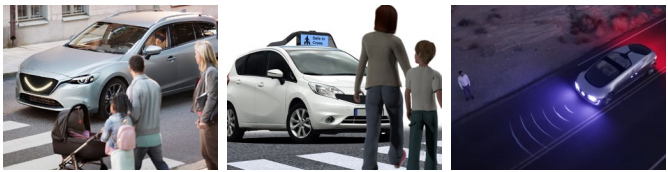
Fig. 1. Some examples of eHMI on vehicles.

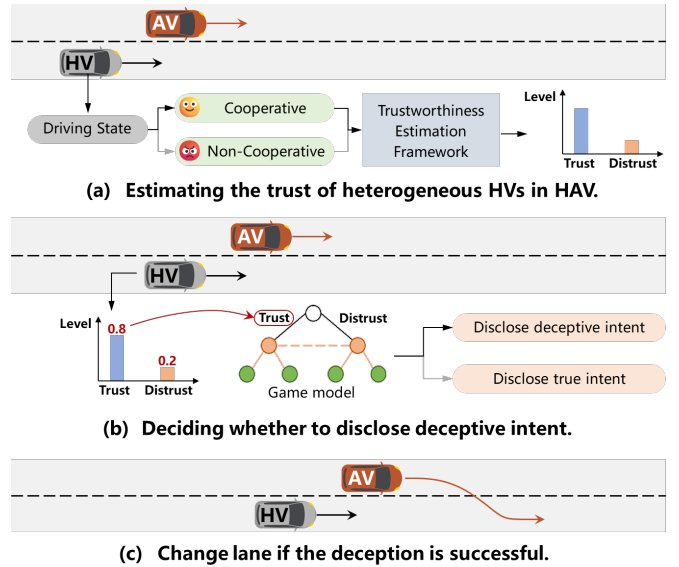
Fig. 2. Illustration of HAV using eHMI for benevolent deception in a lane change scenario. (a) Infers HV driver type to estimate trust. (b) Uses trust in game-theoretic model to assess deception benefit. (c) If successful, HAV changes lane with improved safety and efficiency for both.

The paper is structured as follows: Section 2 reviewed the research on alignment in autonomous driving and trust modeling in human-machine interaction. Section 3 outlined the methodology, including trust estimation and human-machine interaction modeling. Section 4 presented numerical experiments on data processing and model calibration. Section 5 explored the effect of benevolent deception through a case study. Section 6 concluded and discussed the future work.

II. LITERATURE REVIEW

### A. Alignment Problem of Autonomous Driving

The alignment problem refers to the challenge of ensuring that artificial intelligence systems act in ways consistent with human values. While this issue has been widely discussed in the context of large language models and general-purpose agents [3], it has received comparatively little attention in the domain of autonomous driving. Unlike conversational systems, autonomous vehicles operate in safety-critical environments where decisions carry ethical and legal consequences, creating unique demands for alignment.

Previous studies in this area have often focused on extreme moral dilemmas such as the "trolley problem", but lack a broader view of how autonomous vehicles should align with the dynamic moral expectations of society at large. For example, Shah et al. [9] examined public preferences in unavoidable crash scenarios and found significant cultural variation in how harm should be distributed. This discourse finds renewed urgency, also reflected in the IEEE Ethically Aligned Design framework [10], through its call for holistic, adaptive ethical architectures for everyday interactions.

Within this context, the concept of "benevolent deception" offers a meaningful lens through which to examine alignment challenges. It reflects how flexible ethical deviations can support altruistic goals. Bakir et al. [11] argue that simplistic critiques of deception in human-machine interaction should be overcome in favor of a more holistic and empathetic perspective, enabling clearer distinctions between harmful and benign forms of deception. However, current research rarely addresses the potential impact of benevolent deception on external human-machine interaction, especially as HAVs become widely equipped with eHMI and human-machine interactions grow more frequent and complex. Our work built on these discussions by exploring the ethical and practical implications of benevolent deception in autonomous driving.

*B. Human Trust Estimation in Human-Machine Interaction*

Trust plays a pivotal role in the alignment of autonomous systems with human expectations. In human-machine interaction, trust governs the degree of cooperation and the willingness of humans to adapt their behavior in response to autonomous agents [12]. Trust is context-sensitive and not static, evolving over time through observable cues and interaction outcomes.

Current approaches to trust estimation include Bayesian inference, subjective logic and Dempster-Shafer theory, which allow trust to be dynamically estimated and updated under uncertainty [13]. In autonomous driving, such estimation is essential for anticipating whether human drivers will accept or resist HAVs' actions, especially during negotiation-intensive maneuvers like lane changes. Notably, trust is inherently subjective and often difficult to infer objectively. However, social psychology offers a viable approach through observable external responses, enabling reason-based trust estimation, which differs from feeling-based trust shaped by personal experience or belief [14].

Game-theoretic models have become increasingly prevalent for modeling interactive decision-making in autonomous driving [15], [16]. Besides, Bayesian inference provides a natural and rigorous approach for modeling trust. While Bayes' rule has been widely used within game-theoretic frameworks to estimate uncertainty in driver styles and behaviors [17], it has seldom been applied for trust estimation in these contexts. Our work proposed a trust estimation framework integrated into a human-machine interaction game model, where belief updating based on observed behaviors enables human trust to be modeled as a dynamic belief over human intent, facilitating adaptive reasoning under uncertainty.

## III. METHODOLOGY

Among the vehicle interaction modeling approaches, game theory-based methods are adopted in this study due to their explicit behavioral interaction mechanisms and interpretable payoff functions. To investigate the impact of deceptive information disclosed by eHMI on human-machine interaction, this work innovatively incorporates a dynamic trust estimation mechanism into the game-theoretic framework, enabling the quantification of human trust levels and willingness to cooperate.

In many countries, the main liability often falls on the lane-changing vehicle, and this is exacerbated in cases of deception. Therefore, this paper considers the HAV as the lane-changing vehicle and the human vehicle (HV) as the rear vehicle in the target lane.

*A. Human Trust Estimation*

Trust is a subjective and emotional concept, making it difficult to quantify. It can both result from cooperation and foster more cooperation. To provide a more objective measure of human trust for HAV decision-making eHMI, we inferred human trust by observing driver behaviors, such as acceleration and deceleration, based on attribution theory from social psychology [18]. Human trust states can be quantified by their willingness to cooperate with HAVs in lane-changing tasks, as shown in (1), (2).

$$\tau_t^{hv} = P\left(type(HV) = Coop. \mid a_1^{hv}, a_2^{hv}, \ldots, a_{t-2}^{hv}\right) \quad (1)$$

$$type(HV) \in \{Coop., Non-Coop.\} \quad (2)$$

$\tau_t^{hv}$ denotes human trust as the probability of cooperation with HAVs at $t$ moment, which ranges from 0 to 1. $a^{hv}$ represents the actions of human drivers, including acceleration ($v^+$), deceleration ($v^-$) and maintaining speed ($v^m$). $type(HV)$ represents the type of HV and is classified as cooperative (*Coop.*) or non-cooperative (*Non-Coop.*) based on HV's responses to information disclosed by the eHMI. Assuming that a driver's objective behavior adequately reflects their psychological state, we considered the driver to be cooperative if the proportion of deceleration behaviors in the interactions exceeds a threshold $\rho_c$ (it is set to 50% in this study). Recent actions are weighted heavily to improve inference accuracy, as in (3).

$$type(a^{hv}) = \begin{cases} Coop., & \dfrac{\sum_{t=1}^{T} e^{-(T-t)} \cdot \mathbb{I}_{a_t^{hv} < -1\,m/s^2}}{\sum_{t=1}^{T} e^{-(T-t)}} \geq \rho_c \\ Non-Coop., & \dfrac{\sum_{t=1}^{T} e^{-(T-t)} \cdot \mathbb{I}_{a_t^{hv} < -1\,m/s^2}}{\sum_{t=1}^{T} e^{-(T-t)}} < \rho_c \end{cases} \quad (3)$$

where $\mathbb{I}_{a_t^{hv} < -1\,m/s^2}$ is the indicator function, equal to 1 if the condition holds, 0 otherwise.

Human trust in HAVs is dynamic and requires dynamic estimation. We adopted a bayesian inference framework, where trust is iteratively refined based on prior beliefs and newly observed data. Under eHMI influence, human drivers respond through their actions, with cooperative behavior indicating trust. Trust is then estimated via (4).

$$\tau_t^{hv}\left(c \mid a_{t-1}^{hv}\right) = \dfrac{\tau_{t-1}^{hv}(c) P\left(a_{t-1}^{hv} \mid c\right)}{\tau_{t-1}^{hv}(nc) P\left(a_{t-1}^{hv} \mid nc\right) + \tau_{t-1}^{hv}(c) P\left(a_{t-1}^{hv} \mid c\right)} \quad (4)$$

$\tau^{hv}(c)$ and $\tau^{hv}(nc)$ denotes HAV's prior belief in the HV being cooperative or non-cooperative and ranges from 0 to 1. Moreover, due to the independence of the two HV types, it follows that $\tau_{t-1}^{hv}(nc) = 1 - \tau_{t-1}^{hv}(c)$. The initial belief $\tau_{t=0}^{hv}(c)$ can be calculated based on historical interaction behavior data of HVs. $P(a_{t-1}^{hv}|c)$ and $P(a_{t-1}^{hv}|nc)$ are probabilities over actions, given the HV is cooperative or uncooperative, respectively. They are estimated from HV's behavior, as shown in (5), (6).

$$P(a_{t-1}^{hv}|c) = \frac{n_{1\sim t-1}(a_{t-1}^{hv})}{t-1}, \quad type(HV) = Coop. \quad (5)$$

$$P(a_{t-1}^{hv}|nc) = \frac{n_{1\sim t-1}(a_{t-1}^{hv})}{t-1}, \quad type(HV) = Non-Coop. \quad (6)$$

$n_{1\sim t-1}$ responses to the frequency of each action taken by the HV from 1 to $t-1$ moment.

*B. Game Formulation for Lane Changing*

While game theory is widely used for modeling interactive lane-changing behavior, the impact of eHMI is largely neglected. However, the information disclosed via eHMIs inevitably affects human-machine interactions, particularly when deceptive information is involved. Our work aims to develop an incomplete information game model incorporating a dynamic trust evaluation mechanism to guide the decision-making of HAVs and HVs.

*1) Strategy Sets and Game Structure*

Lane changes are classified by intent into Mandatory Lane Change (MLC) and Discretionary Lane Change (DLC). In MLC, as shown in Fig. 3(a), the HAV in the inner lane must exit via an off-ramp, prioritizing safety even if it requires deceleration or waiting. In DLC, as shown in Fig. 3(b), the HAV aims to improve efficiency by changing lanes when the adjacent lane offers better acceleration opportunities and safe gaps are available. Our work focused on the most influential interacting HV, which is the first rear HV in the target lane.

Based on the analysis of the lane-changing scenarios, the HAV controls the eHMI to disclose information, with strategies including communicating **lane-changing intent ($LC$)** and **yielding intent ($Yield$)**. The strategies of the HV as the rear vehicle in the target lane are **accelerating**, **decelerating** and **maintaining current speed**. As HAVs and HVs cannot establish a cooperative agreement in advance and both aim to optimize driving efficiency, the decision-making process during lane changing is regarded as a non-cooperative game. Furthermore, with eHMI intervention, HVs may decide after observing the information disclosed by HAVs, aligning with the temporally sequential characteristics of dynamic games. Considering computational complexity, the decision process at each moment is modeled using a single-stage dynamic game.

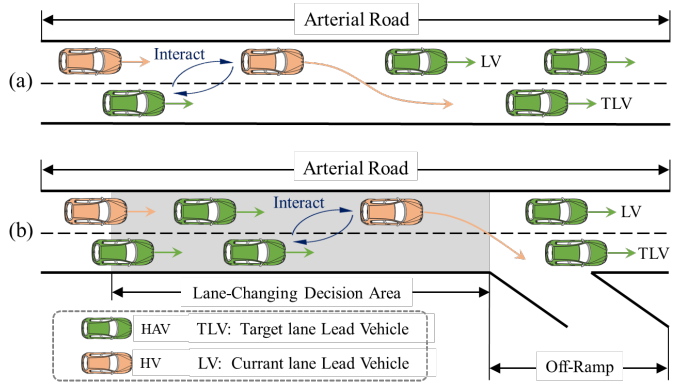

Fig. 3. Illustration of Lane Change Scenarios. (a) Discretionary Lane Change (DLC) prioritizing efficiency while ensuring safety conditions; (b) Mandatory Lane Change (MLC) prioritizing safety, sacrificing efficiency to ensure successful lane change.

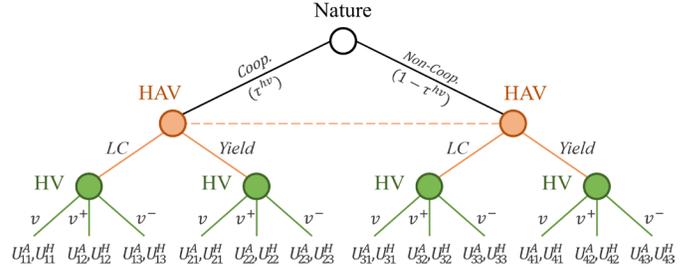

Fig. 4. Tree diagram of the incomplete information dynamic game. HAV discloses intent through eHMI as the first mover, while HV selects its strategy based on the disclosed information.

The Harsanyi transformation [19] introduces a virtual player "Nature" to model uncertainty in human trust, converting the incomplete information game to a complete but imperfect information game. Specifically, Nature first selects the type of HV as cooperative or uncooperative based on the HAV's belief. Then HAV decides what information to disclose through eHMI, and the HV makes a decision based on this information. The game tree is shown in Fig. 4.

The HAV initially holds a prior belief $\tau_{t=0}^{hv}$ regarding the type of HV. As interactions proceed, HAV observes HV's strategies and updates the posterior probability $\tau_t^{hv}(c|a_{t-1}^{hv})$ accordingly using Bayes' rule.

*2) Payoff Formulations*

We formulated the payoff functions of HAV and HV under MLC and DLC scenarios, considering both safety and efficiency. Let $U^A$ and $U^H$ denote the payoffs of the HAV and HV, respectively. The payoffs under different strategies are shown in Table I.

HAVs prioritize safety in terms of MLC, and efficiency is omitted from the payoff. While in terms of DLC, both HAV and HV consider safety and efficiency. Additionally, the HAV's payoff is incorporated into the cooperative HV's payoff. Thus, the payoffs of HAV and HV under the strategy sets $(i, j)$ can be expressed as follows:

$$U_{ij}^A = w_s \cdot S_{ij} + w_e \cdot E_{ij} \cdot (1 - \mathbb{I}_{MLC}) \quad (7)$$

TABLE I. THE PAYOFF LIST OF GAME

| HV | Cooperative HV ($\tau^{hv}$) | | Uncooperative HV ($1-\tau^{hv}$) | |
|---|---|---|---|---|
| | **LC** | **Yield** | **LC** | **Yield** |
| $v$ | $U_{11}^A, U_{11}^H$ | $U_{21}^A, U_{21}^H$ | $U_{31}^A, U_{31}^H$ | $U_{41}^A, U_{41}^H$ |
| $v^+$ | $U_{12}^A, U_{12}^H$ | $U_{22}^A, U_{22}^H$ | $U_{32}^A, U_{32}^H$ | $U_{42}^A, U_{42}^H$ |
| $v^-$ | $U_{13}^A, U_{13}^H$ | $U_{23}^A, U_{23}^H$ | $U_{33}^A, U_{33}^H$ | $U_{43}^A, U_{43}^H$ |

$$U_{ij}^{H} = \begin{cases} \tau^{hv} \cdot (w_s \cdot S_{ij} + w_e \cdot E_{ij} + w_a U_{ij}^{A}), & i \in \{1, 2\} \\ (1-\tau^{hv}) \cdot (w_s \cdot S_{ij} + w_e \cdot E_{ij}), & i \in \{3, 4\} \end{cases} \quad (8)$$

where $S$ denotes the payoff of safety, $E$ denotes the payoff of efficiency. $\mathbb{I}_{MLC}$ is an indicator function that takes the value of 1 if the scenario involves MLC. $w_e$ and $w_s$ represent the weights assigned to safety and efficiency, respectively. $w_a$ denotes the weight reflecting the synergistic consideration of the payoff of HAV.

The time-to-collision (TTC) is used as the indicator in $S$. For HAV lane changes, safe distances to the TLV (lead vehicle in the target lane) and rear HV in the target lane are required. When HAV yields, the distance to LV (lead vehicle in the current lane) in the current lane is considered. Conversely, HVs evaluate safety with respect to the HAV during its lane change, or to the TLV when the HAV yields. For each $X \in \{HAV, HV\}$, under strategic set $(i,j)$, the formulation of $S_{ij}$ is given as follows:

$$S_{ij}^{X} = \exp\left(-1 \Big/ \min_{B \in \mathcal{R}_{ij}^{X}} \left(\frac{x_B - x_X}{v_X - v_B}\right) + \varepsilon\right) \quad (9)$$

$$\mathcal{R}_{ij}^{HAV} = \begin{cases} \{TLV, HV\}, & i \in \{1,3\} \\ \{LV\}, & i \in \{2,4\} \end{cases} \quad (10)$$

$$\mathcal{R}_{ij}^{HV} = \begin{cases} \{HAV\}, & i \in \{1,3\} \\ \{TLV\}, & i \in \{1,3\} \end{cases} \quad (11)$$

where $x$ denotes the position coordinate of the vehicle along the driving direction $(m)$. $v$ denotes the vehicle's speed $(m/s)$. $\varepsilon$ is a small constant to prevent division by zero.

Efficiency payoff primarily quantifies the potential speed gain during car-following, increasing with the lead vehicle's speed, acceleration, and headway. For each $X \in \{HAV, HV\}$, under strategic set $(i,j)$, the formulation of $E_{ij}$ is shown in (12):

$$E_{ij}^{X} = v_{\mathcal{F}_{ij}^{X}} - v_X + a_{\mathcal{F}_{ij}^{X}} \cdot \left(x_{\mathcal{F}_{ij}^{X}} - x_X\right) \quad (12)$$

$$\mathcal{F}_{ij}^{HAV} = \begin{cases} TLV, & i \in \{1,3\} \\ LV, & i \in \{2,4\} \end{cases} \quad (13)$$

$$\mathcal{F}_{ij}^{HV} = \begin{cases} HAV, & i \in \{1,3\} \\ TLV, & i \in \{2,4\} \end{cases} \quad (14)$$

where $a$ denotes the acceleration $(m/s^2)$ required for HV to execute the strategy.

*3) Nash Equilibrium Settings*

Perfect Bayesian Equilibrium (PBE) is employed to solve the game of imperfect information. PBE yields a set of strategies that are sequentially rational given the HAV's beliefs, while the beliefs are continuously updated according to Bayes' rule. Since HAV moves first, we updated its belief based on HV's action in the previous round of the game. The probability of natural selection of HV's type is determined by HAV's belief in the previous round. Let $\theta = \{\theta^c, \theta^{nc}\}$ represents the type of HV, which is chosen by Nature. $a^{hav} = \{LC, Yield\}$ and $a^{hv} = \{v, v^+, v^-\}$ represent the strategies selected by HAV and HV. $u_{HAV}$ and $u_{HV}$ denote the payoffs of HAV and HV. The PBE must satisfy the following conditions shown in (15, 16, 17).

$$\tau_t^{hv}(\theta \mid a_{t-1}^{hv}) = \frac{\tau_{t-1}^{hv}(\theta) P(a_{t-1}^{hv} \mid \theta)}{\tau_{t-1}^{hv}(\theta^{nc}) P(a_{t-1}^{hv} \mid \theta^{nc}) + \tau_{t-1}^{hv}(\theta^c) P(a_{t-1}^{hv} \mid \theta^c)} \quad (15)$$

$$\forall \theta, \quad \sigma_{HAV}^{*}(\cdot \mid \theta) \in \arg\max_{a^{hav}} \sum_{\theta_i \in \theta} \tau_{hv}(\theta_i) \cdot u_{HAV}(a^{hav}, \theta) \quad (16)$$

$$\forall a^{hav}, \quad \sigma_{HV}^{*}(\cdot \mid a^{hav}, \theta_i) \in \arg\max_{a^{hv}} \sum_{a_i \in a^{hv}} u_{HAV}(a^{hav}, a_i, \theta_i) \quad (17)$$

IV. NUMERICAL EXPERIMENT

*A. Data Processing*

We conducted a cross-cultural case study using the HighD [20] and exiD [21] datasets from Germany and the MAGIC dataset [22] from China. Vehicle speeds on German highways are significantly higher than those in China. Furthermore, a prior study [23] found that Chinese drivers showed high acceptance of HAV decisions across scenarios, while US and German drivers preferred overtaking in low-hindrance cases and rejected decisions causing high hindrance. We extracted 1,053 DLC and 992 MLC trajectories from HighD&exiD, and 816 DLC and 753 MLC trajectories from MAGIC dataset. The bird's-eye view of the road segments and examples of observed lane-change trajectories are shown in Fig. 5.

Since trajectories reflect interaction outcomes rather than intent, we performed frame-by-frame analysis of vehicle motion. Driving strategies and intentions were inferred based on lateral and longitudinal acceleration patterns.

*B. Model calibration*

The calibration framework proposed by Liu et al. [24] is adopted to calibrate our model's parameters. Specifically, the upper level solves a nonlinear optimization problem to minimize the gap between predicted and observed actions. Due to the complexity of strategies and payoffs, gradient descent often fails to converge optimally. Thus, a genetic algorithm is employed for more robust optimization. The lower level solves for the PBE, with predicted strategies passed to the upper level, which in turn updates the parameters.

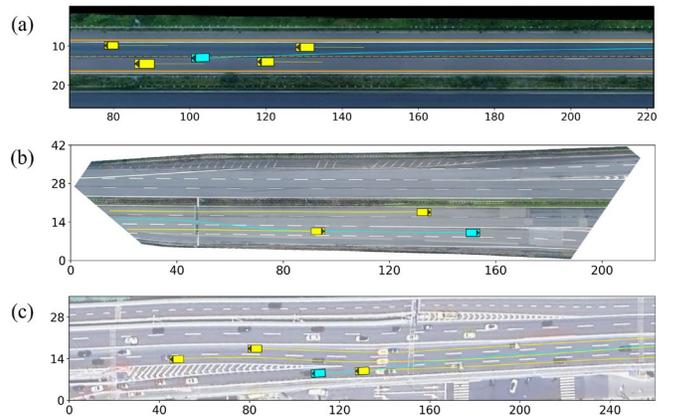

Fig. 5. Examples of observed lane-change trajectories from different datasets. (a) HighD dataset featuring discretionary lane changes (DLC) only. (b) exiD dataset containing both DLC and mandatory lane changes (MLC). (c) MAGIC dataset containing both DLC and MLC.

TABLE II. RESULTS OF MODEL CALIBRATION

| Dataset | Scen. | Para. | Result | Para. | Result | Para. | Result |
|---|---|---|---|---|---|---|---|
| HighD& exiD | MLC | $w_e$ | 0.292 | $w_s$ | 0.734 | $w_a$ | 0.429 |
| | DLC | $w_e$ | 0.651 | $w_s$ | 0.677 | $w_a$ | 0.406 |
| MAGIC | MLC | $w_e$ | 0.433 | $w_s$ | 0.518 | $w_a$ | 0.806 |
| | DLC | $w_e$ | 0.724 | $w_s$ | 0.529 | $w_a$ | 0.751 |

Prior to model calibration, we estimated the prior trust of HVs using the clustering approach. Under the assumption that drivers exhibiting conservative behavior are more inclined to cooperate, we approximated the prior trust level by the proportion of conservative driving styles within each dataset. The resulting estimates were 0.228 for German drivers and 0.329 for Chinese drivers, aligning with existing hypotheses regarding cross-cultural differences in cooperative driving tendencies.

80% of the trajectory data was used for model calibration, with the remaining 20% reserved for validation. The True Positive Rate (TPR) is used as an indicator to assess model predictability. Prediction accuracy reached 93.3% on the HighD&ExiD datasets and 90.7% on the MAGIC dataset. Parameter calibration results are shown in Table II.

## V. THE CASE STUDY AND RESULT ANALYSIS

### A. eHMI Information Disclosure Mechanism

In the absence of eHMI, game strategies directly correspond to actual behavior. With eHMI, however, disclosed intent may differ from intended strategy. We defined "deception" as a mismatch between the HAV's disclosed intent and its intended action. A "benevolent deception" occurs when a HAV discloses a deceptive intent that strengthens HVs' willingness to yield, reducing collision risk and improving efficiency. This works by shortening hesitation and trial-and-error interactions, ensuring both vehicles adopt clear, cooperative strategies.

As illustrated in Fig. 7, benevolent deception allows the HAV to influence HV behavior and gain strategic advantage. In Fig. 7(a), the HAV discloses its true lane-change intent, and successfully changes lanes after the HV yields. In Fig. 7(b), the HAV's expected strategy is to change lanes, but it discloses a deceptive yielding intention. This causes the HV into yielding, resulting in a Nash equilibrium where both parties yield. As a result, the HAV can create a larger gap and change lanes with minimal negotiation, improving efficiency while maintaining safety. A specific process of deception is:

1) **Deceptive information release**: The HAV discloses a deceptive yield intention through eHMI.
2) **HAV select strategy:** When the Nash equilibrium state results in both vehicles yielding, the deception is considered successful. And HAV will change lanes.

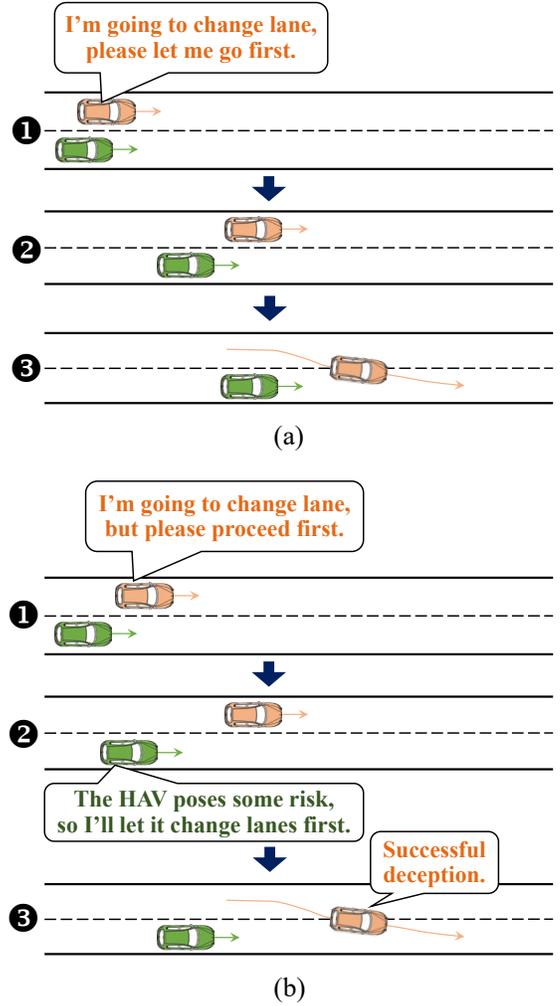

Fig. 7. Illustration of the HAV lane-change process with and without benevolent deception. (a) HAV discloses true lane-change intent and successfully changes lanes after HV yields, (b) HAV discloses deceptive yielding intent, causing HV to yield and allowing the HAV to change lanes with a larger gap and minimal negotiation.

3) **Identifying HV's response**: The HAV continuously observes HV's acceleration to infer its intent and calculate its trust level.
4) **Trust protection mechanism**: If the HV's trust declines over interactions and falls below a defined threshold, the HAV ceases deception and reveals its true intent.

Within this information disclosure framework, the HAV dynamically estimates the HV's trust level and integrates it into the game-theoretic decision-making process. A predefined trust threshold is used to determine the termination of deception, thereby safeguarding human drivers' trust throughout interactions.

### B. Efficiency improvement performance

Lane change time is selected as the indicator for interaction efficiency. It refers to the time interval from the initiation to the completion of a single lateral movement by the HAV. A shorter lane change time indicates higher efficiency, as it signifies the

HAV's ability to confidently and swiftly complete the lane change without repeated negotiations.

Fig. 8 shows the lane-change time distributions with and without the implementation of deception under two datasets. As shown in the figure, disclosing deceptive intention, prompts the HV to yield quickly, facilitating a faster lane change. The results also confirm that deception is more successful in the Chinese scenario, where HVs are more likely to cooperate. In the Chinese scenario where the HAV's expected strategy was Yield, deceptive behavior led to the greatest improvement in lane-changing efficiency, with 59.4% of cases showing reduced lane-change time. Furthermore, we found that when the HAV expects to yield but signals a lane change, the deception success rate is 56.7%, compared to 21.3% when it expects to change lanes but signals yielding.

### C. Safety improvement performance

The Time Difference to Collision (TDTC) is used to evaluate the safety of lane changes. It indicates the time until vehicles reach the conflict point with constant speed and direction. A smaller TDTC indicates a more severe collision risk. Fig. 9 shows the lane-change time distributions with and without the implementation of deception under two datasets.

Due to the large TDTC when speed differences are small, which makes visualization difficult, we plotted the inverse of TDTC. It can be observed that disclosing deceptive intentions allows for safer interactions between the HAV and HV. With a Yield expectation, deception most significantly enhanced safety, increasing TDTC in 43.6% of German cases and 52.7% of Chinese cases.

### D. Discussions of Deception

In this study, we set the trust collapse threshold for HV at 50% trust loss, at which point deception is terminated before collapse occurs. This threshold-based strategy reflects the core tension between goal alignment, which focuses on pursuing traffic efficiency and safety, and value alignment, which emphasizes maintaining ethical integrity and human trust. The results of the case studies indicate that the proportion of HV trust collapse in the German scenario is 36.9%, whereas in the Chinese scenario, it is only 13.2%. Fig. 10(a) and Fig. 10(b) depict the dynamics of HVs' trust in HAVs during trust recovery and trust collapse scenarios, respectively. Trust is represented as the probability of HV cooperation with HAV. In Fig. 10(a), trust declines to a minimum of approximately 0.39 before recovering to around 0.84, though it does not reach the initial level. In contrast, Fig. 10(b) shows a trust collapse. Although the HAV ceases the deceptive behavior, trust continues to decline, reaching as low as 0.09, and fails to recover. One contributing factor is the aggressive driving style of the HV driver in this scenario, which correlates with a lower initial trust level in the HAV. As a result, once trust is damaged, it is much harder to regain. This suggests that in European scenarios, where drivers are more self-focused and less cooperative, trust is more likely to collapse and much harder to recover once broken. These findings prompt further consideration of whether it is permissible to allow HAVs to employ benevolent deception in real-time. While such benevolent deception could theoretically enhance interaction efficiency and safety, the collapse of trust could lead to broader, more long-term, and unpredictable detrimental effects.

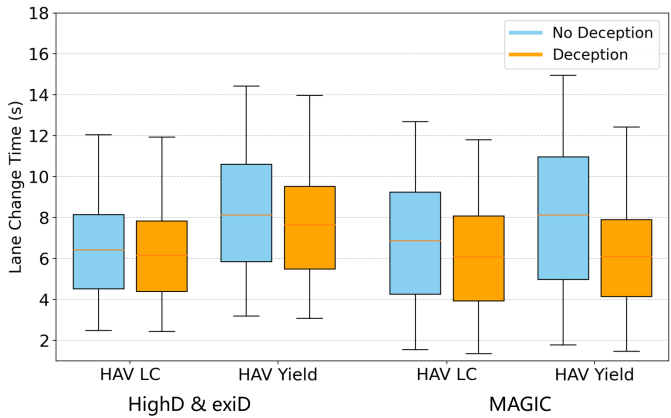

Fig. 8. Lane change time distribution under varying information disclosure conditions. In all cases, lane change time is reduced due to enhanced HV yielding willingness, allowing quicker interactions without repeated negotiations.

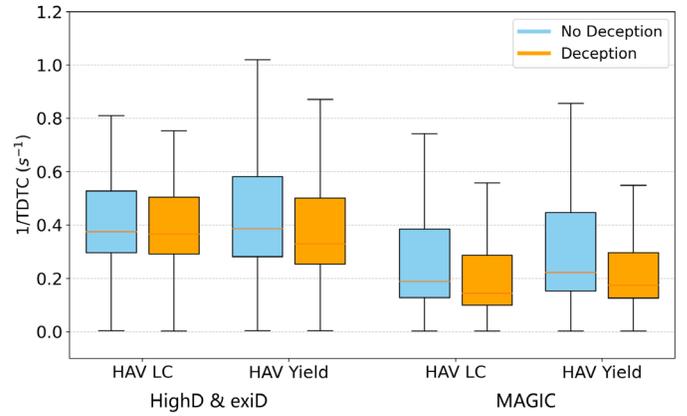

Fig. 9. TDTC under varying information disclosure conditions. In all cases, TDTC is reduced due to enhanced HV yielding willingness and more decisive deceleration. Deception leads to a decrease in collision risk, allowing interactions to conclude more safely.

Nevertheless, we remained optimistic about aligning ADS with adaptive ethical principles to enhance HV's willingness to cooperate. The dynamic trust estimation framework introduced in our methodology plays a crucial role in this alignment process, allowing HAVs to personalize their interaction strategies based on observed human behavior patterns. By setting a higher trust collapse protection threshold and utilizing advanced driver behavior recognition to predict and select cooperative drivers, deception could become more feasible and harder to detect. While benevolent deception may resemble human strategies or ethically controversial actions intended for greater good, its deployment by machines raises unresolved questions about interpretability, accountability, and ethical legitimacy. Especially, at present, the intelligence of ADS and the maturity of relevant regulations are insufficient to fully support perfect deception.

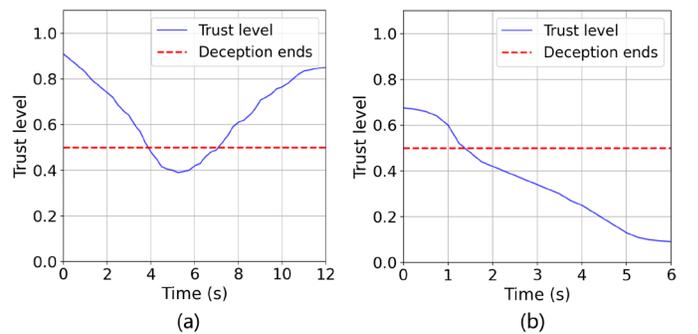

Fig. 10. Cases of HV Trust Dynamics. (a) Under deception, HV's trust decreases from 0.91 to 0.39. Once it falls below 0.5, HAV terminates the deception, after which HV's trust recovers to 0.84. (b) Under deception, HV's trust drops from 0.89 to 0.09, resulting in a complete collapse with no chance of recovery.

## VI. CONCLUSION

In environments where HAVs and HVs coexist, ensuring ADS decisions align with human values is crucial. This study proposed a framework examining the impact of HAVs using eHMI to disclose benevolent deceptive intentions on lane-changing interactions, incorporating dynamic trust estimation, incomplete information games, model calibration, and case studies. And case studies showed that disclosing deceptive information through eHMI can improve efficiency and safety for both HAV and HV.

However, it is important to acknowledge the potential drawbacks and ethical concerns associated with this approach. While benevolent deception may facilitate smoother interactions in the short term, it risks undermining long-term trust if human drivers perceive the system as manipulative or inconsistent. Repeated exposure to deceptive cues, even if well-intentioned, could desensitize human drivers or lead to misinterpretation, especially in edge cases or high-stakes scenarios.

In addition, there are several limitations in the current study: 1) trust from human drivers is inferred from observable behavior, which may not fully capture their underlying psychological states or perceptions of fairness; 2) the lack of real-world simulation or on-road validation limits the generalizability and applicability of our findings; 3) the study focuses only on short-term trust effects, without addressing how repeated interactions may influence long-term trust development. In conclusion, while our work offered valuable insights into improving human-machine interactions through benevolent deception, further research and real-world validation are needed to refine its practical applications.